\input{aipcheck}

\documentclass[
    ,final            % use final for the camera ready runs
  ]
  {aipproc}

\layoutstyle{6x9}

\begin{document}

\title{Dynamical Model for Meson Production off Nucleon
and Application to Neutrino-Nucleus Reactions}

\classification{25.30.-c,25.30.Pt,13.60.Le}
% 13.60.Le : Meson production by photons and leptons, 
% 25.30.-c : Electron-induced nuclear reactions, Inelastic scattering in
% lepton-induced reactions, lepton-induced nuclear reactions, electron
% in nuclear reactions
% 25.30.Pt : Neutrinos in nuclear scattering, neutrino-induced Nuclear
% reactions, neutrino-nucleus scattering 
%
\keywords      {Neutrino-nucleus reaction, Neutrino-induced pion production}

\author{Satoshi X. Nakamura}{
  address={Excited Baryon Analysis Center (EBAC)\\ 
Thomas Jefferson National Accelerator Facility, Newport News, Virginia 23606, USA}
}

\begin{abstract}
I explain the Sato-Lee (SL) model and its extension to
the neutrino-induced pion production off the nucleon. 
Then I discuss applications of the SL model to incoherent and coherent pion
productions in the neutrino-nucleus scattering.
I mention a further extension of this approach with
a dynamical coupled-channels model developed in
Excited Baryon Analysis Center of JLab.
\end{abstract}

\maketitle

%%%%%%%%%%%%%%%%%%%%%%%%%%%%%%%%%%%%%%%%%%%%
%% MAINMATTER
%%%%%%%%%%%%%%%%%%%%%%%%%%%%%%%%%%%%%%%%%%%%

\section{Introduction}

Neutrino oscillation experiments have been actively conducted in the
last decade, and will be so in the forthcoming decade.
Because those experiments detect the neutrino through the
neutrino-nucleus ($\nu$-$A$) scattering, understanding of the $\nu$-$A$
scattering is a prerequisite for a successful interpretation of data.
Some of the experiments measure the neutrino in the energy region of
sub- and few-GeV where dominant processes are quasi-elastic nucleon
knockout (QE) and single pion ($1\pi$) production via the $\Delta$-excitation.
For QE, the elementary amplitude is reasonably well-known, and the
challenge is to incorporate the nuclear correlation in the initial state
and the final state interaction. 
Although there has been a reasonable success in describing
QE in the electron-nucleus scattering, it was reported that
the same framework does not work well for the $\nu$-$A$ scattering~\cite{benhar1}.
From here, I focus on the $1\pi$ productions in $\nu$-$A$ scattering in
the $\Delta$ region, 
which constitute the dominant background in the neutrino oscillation experiments.
In addition to the difficult problem of the nuclear effects,
relevant elementary amplitudes, or
dynamical models which generate them, also have to be carefully studied.
Those dynamical models are developed through a careful
analysis of data for electroweak $1\pi$ production off the nucleon.
Actually, there have been active developments of such dynamical models,
motivated by extensive experiments of photo- and electro
meson-productions in the resonance region. 
These experiments aim to test resonance properties predicted by
QCD-inspired models and Lattice QCD. 
A dynamical model for $1\pi$ production developed in this way provides a
good starting point to study the neutrino-induced $1\pi$ production off
the nucleon, because of the close relation between the weak and the
electromagnetic currents. 
Furthermore, the dynamical model offers a good basis to study $1\pi$
production in the $\nu$-$A$ scattering.

Thus first, I give a brief description of a dynamical model, i.e.,
the Sato-Lee (SL) model~\cite{sl}, for the $1\pi$ photo-production
off the nucleon. (For a fuller discussion, consult Ref.~\cite{sl}.)
Then I discuss the extension of the SL model to the weak sector, done in Ref.~\cite{SUL}.
The elementary amplitudes generated by the SL model has been applied to
pion productions in $\nu$-$A$ scattering.
I describe the work done in Ref.~\cite{sl-1pi} where the authors studied
the quasi-free $\Delta$-excitation followed by the single pion
production.
I also discuss the coherent pion production off a nucleus 
studied with the SL model~\cite{coh}.
Finally, I discuss a possible future development.

\section{Sato-Lee (SL) model}

In the SL model, one starts with a set of phenomenological Lagrangians,
and derive an effective Hamiltonian using a unitary transformation.
The effective Hamiltonian for pion photoproduction can be written as follows:
\begin{eqnarray}
H_{eff} & = & H_0 + v_{\pi N} + v_{\gamma\pi}
 + \Gamma_{\pi N \leftrightarrow \Delta} 
+ \Gamma_{\gamma N \leftrightarrow \Delta},   \label{hamile}
\end{eqnarray}
where $H_0$ is the free Hamiltonian, 
$v_{\pi N}$ and $v_{\gamma\pi}$ are respectively non-resonant 
$\pi N\to \pi N$ and $\gamma N\to \pi N$ potentials, 
and are composed by the Born diagrams, 
$t$-channel $\rho$ and $\omega$ 
exchange terms, and the crossed $\Delta$ term.
Bare vertices for $\pi N\leftrightarrow\Delta$ and $\gamma N\leftrightarrow\Delta$
transitions are respectively denoted by
$\Gamma_{\pi N \leftrightarrow \Delta}$ and
$\Gamma_{\gamma N \leftrightarrow \Delta}$.
With the effective Hamiltonian, we can derive unitary pion photoproduction
amplitude as
\begin{eqnarray}
T_{\gamma\pi}(E)  =  t_{\gamma\pi}(E) + 
\frac{
\bar{\Gamma}_{\Delta \rightarrow \pi N}(E)
\bar{\Gamma}_{\gamma N \rightarrow \Delta}(E)
}
{E - m_\Delta - \Sigma_\Delta(E)} , \label{tmatt}
\end{eqnarray}
where the first (second) term is the nonresonant (resonant) amplitude,
and $E$ is the total energy of the pion and nucleon; $m_\Delta$ is the
$\Delta$ bare mass.
The nonresonant amplitude is calculated by
\begin{eqnarray}
t_{\gamma\pi}(E)=   v_{\gamma\pi} 
+  t_{\pi N}(E)G_{\pi N}(E)v_{\gamma\pi},  \label{tmatg}
\end{eqnarray}
where $G_{\pi N}$ is the $\pi N$ free propagator, 
and $t_{\pi N}$ is obtained by solving Lippmann-Schwinger equation which
includes $v_{\pi N}$.
In Eq.~(\ref{tmatt}), the $\Delta$ vertices are dressed as
\begin{eqnarray}
\bar{\Gamma}_{\gamma N \rightarrow \Delta}(E)  &=&  
   \Gamma_{\gamma N \rightarrow \Delta} + v_{\gamma \pi} G_{\pi N}(E)
\bar{\Gamma}_{\pi N \rightarrow \Delta}(E)  ,  \label{vertg} \\
\bar{\Gamma}_{\Delta \rightarrow \pi N}(E)
 &=& [1+t_{\pi N}(E)G_{\pi N}(E)]\Gamma_{\Delta\rightarrow\pi N} \ , \label{vertp}
\end{eqnarray}
so that dynamical pion cloud effect is taken into account as a
consequence of the unitarity.
The $\Delta$ self-energy in Eq.~(\ref{tmatt}) is given by
\begin{eqnarray}
\Sigma_\Delta(E) = 
 \Gamma_{\pi N\rightarrow \Delta}
G_{\pi N}(E)\bar{\Gamma}_{\Delta \rightarrow \pi N}(E).  \label{self}   
\end{eqnarray}
The $\pi N$ scattering amplitude is calculated similarly. 
Thus one first determine the strong interactions by analyzing $\pi N$
data in the $\Delta$-region. 
Then adjustable parameters relevant to electromagnetic interactions
are fixed by analyzing $\gamma N\to \pi N$ data.
With this approach, the SL model has been shown to give a reasonable
description for the $\pi N$, $\gamma N\to \pi N$~\cite{sl}
and $e N\to e'\pi N$ reactions~\cite{sl2} in the $\Delta$-region.

\section{1$\pi$ production in neutrino-nucleon scattering}

With the SL model for pion photo-(and electro-) production discussed above,
it is straightforward to extend it to the weak sector,
as has been done in Ref.~\cite{SUL}.
One just need to replace the electromagnetic current with the with
$V_\mu-A_\mu$ where $V_\mu$ ($A_\mu$) is the weak vector (axial-vector) current.
The vector current conservation hypothesis tells us that 
the weak vector current is obtained
from the isovector part of the electromagnetic current 
by the isospin rotation. 
Thus the remaining part is the axial-current. 
In Ref.~\cite{SUL}, the authors parametrized the least known axial-vector
$N\Delta$ transition matrix element, more specifically form factors, as
\begin{eqnarray}
 g_{AN\Delta} (Q^2) = 
 g_{AN\Delta} (0) R(Q^2) G_A(Q^2) \ ,
\end{eqnarray}
where $G_A(Q^2)=1/(1+Q^2/M_A^2)^2$ with $M_A=1.02$~GeV. 
The coupling $g_{AN\Delta} (0)$ is related to the nucleon axial coupling
$g_A$ using the nonrelativistic constituent quark model.
The remaining correction factor, $R(Q^2)=(1+aQ^2) e^{-bQ^2}$, is assumed
to be the same as that used for the $\gamma N\to \Delta$ form factor,
and thus is determined by analyzing the pion electroproduction data.

\begin{figure}[t]
\begin{minipage}[t]{75mm}
   \includegraphics[width=70mm]{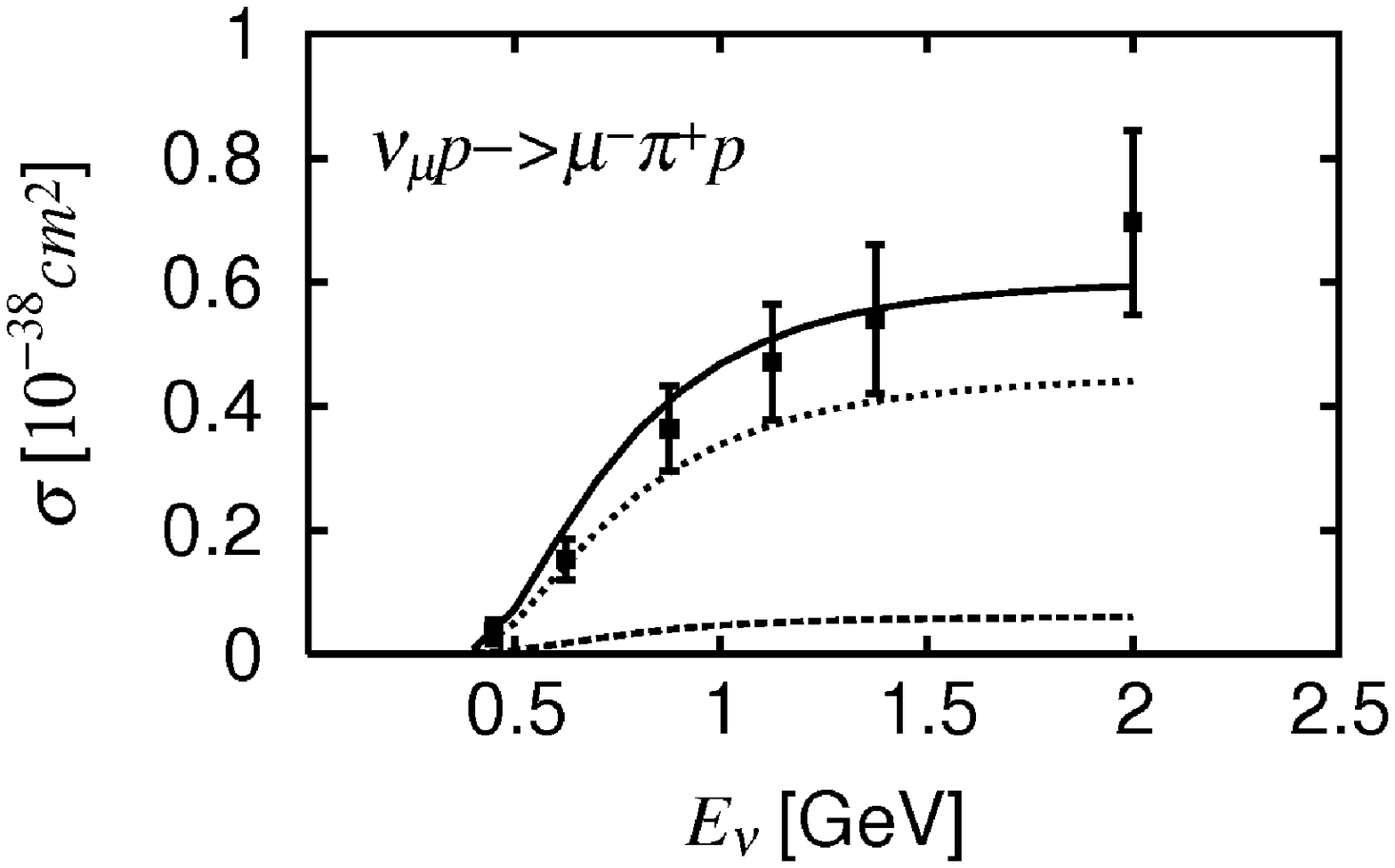}
 \end{minipage}
\begin{minipage}[t]{75mm}
   \includegraphics[width=70mm]{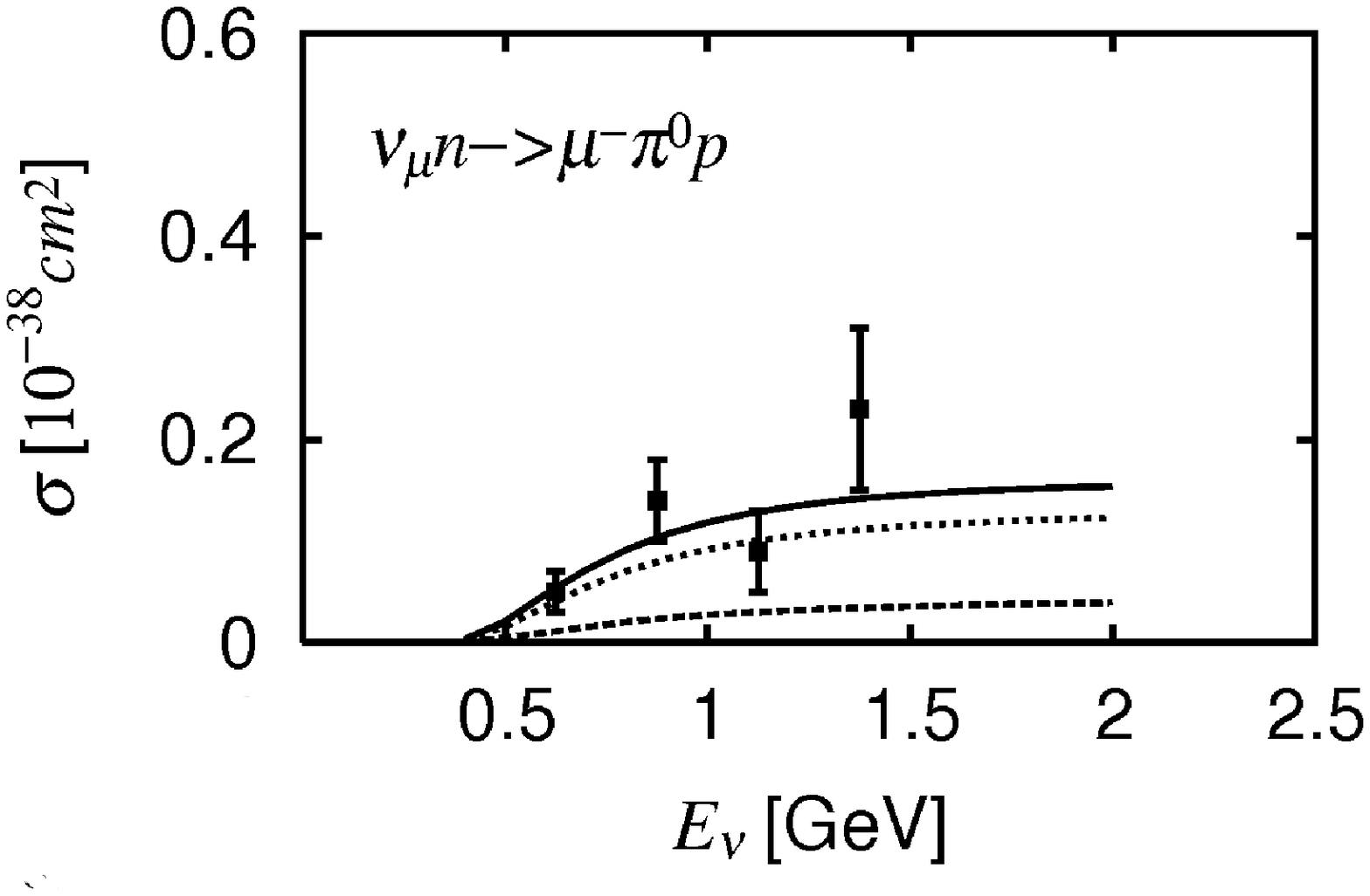}
 \end{minipage}
\caption{\label{fig:nu-tot}
Total cross sections for $\nu_\mu N \to \mu^- \pi N$. For an explanation
 of each curve, see the text. Data are from Ref.~\cite{barish}.}
\end{figure}
The total cross sections predicted by this model is compared with data
in Fig.~\ref{fig:nu-tot}. 
It turns out that
the full calculation (solid curves) shows a good consistency with the data. 
If we turn off the meson cloud effect, then we obtain the dotted curves,
indicating the significant effect.
We further turn off the contribution from the bare $N\Delta$ transition,
then we obtain the dashed curve, showing the non-resonant contribution.
Although the non-resonant contribution is smaller than the resonant one, 
it is still important to get a good agreement with data because it can
interfere with the resonant amplitude.

\section{1$\pi$ production in neutrino-nucleus scattering}

The SL model discussed in the previous section can be applied to
the neutrino-nucleus interaction in the $\Delta$-region, which has been
conducted in Ref.~\cite{sl-1pi} for the $^{12}$C target.
A unique feature of this application is that the SL model treats
resonant and non-resonant mechanism on the same footing so that the
amplitude is unitary, while most previous works considered only
resonant mechanisms. 
The challenge here is to incorporate 
elementary amplitudes generated by the SL model with
various nuclear effects such as:
the nuclear correlation effect in the
initial state; the Pauli blocking on the final nucleons; the final state
interactions (including pion absorption); the medium effect on
the $\Delta$-propagation.
The authors of Ref.~\cite{sl-1pi} considered the initial nuclear
correlation using the spectral function~\cite{benhar}, and the Pauli
blocking using the Fermi gas model.

\begin{figure}[t]
\begin{minipage}[t]{75mm}
   \includegraphics[width=70mm]{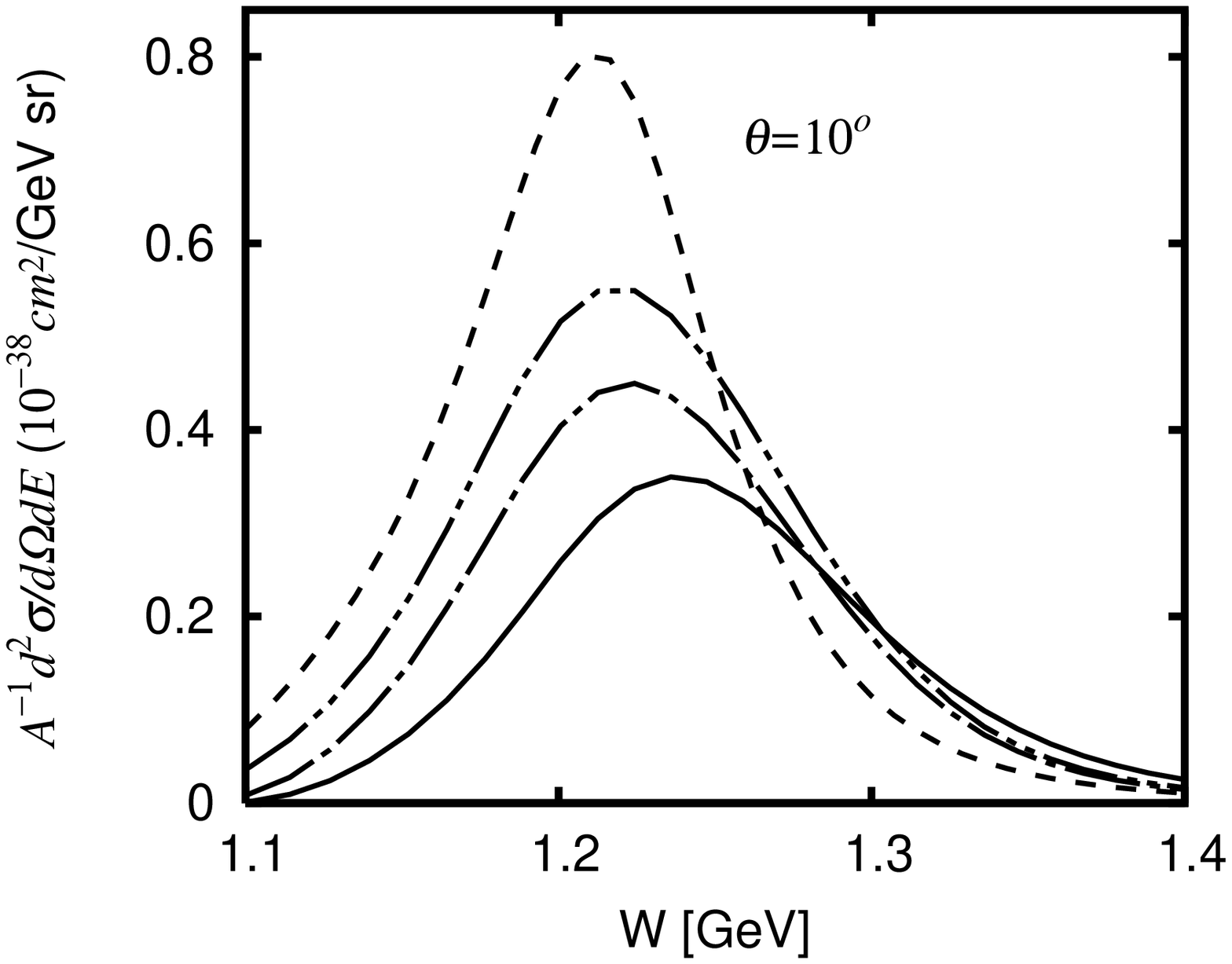}
 \end{minipage}
\begin{minipage}[t]{75mm}
   \includegraphics[width=70mm]{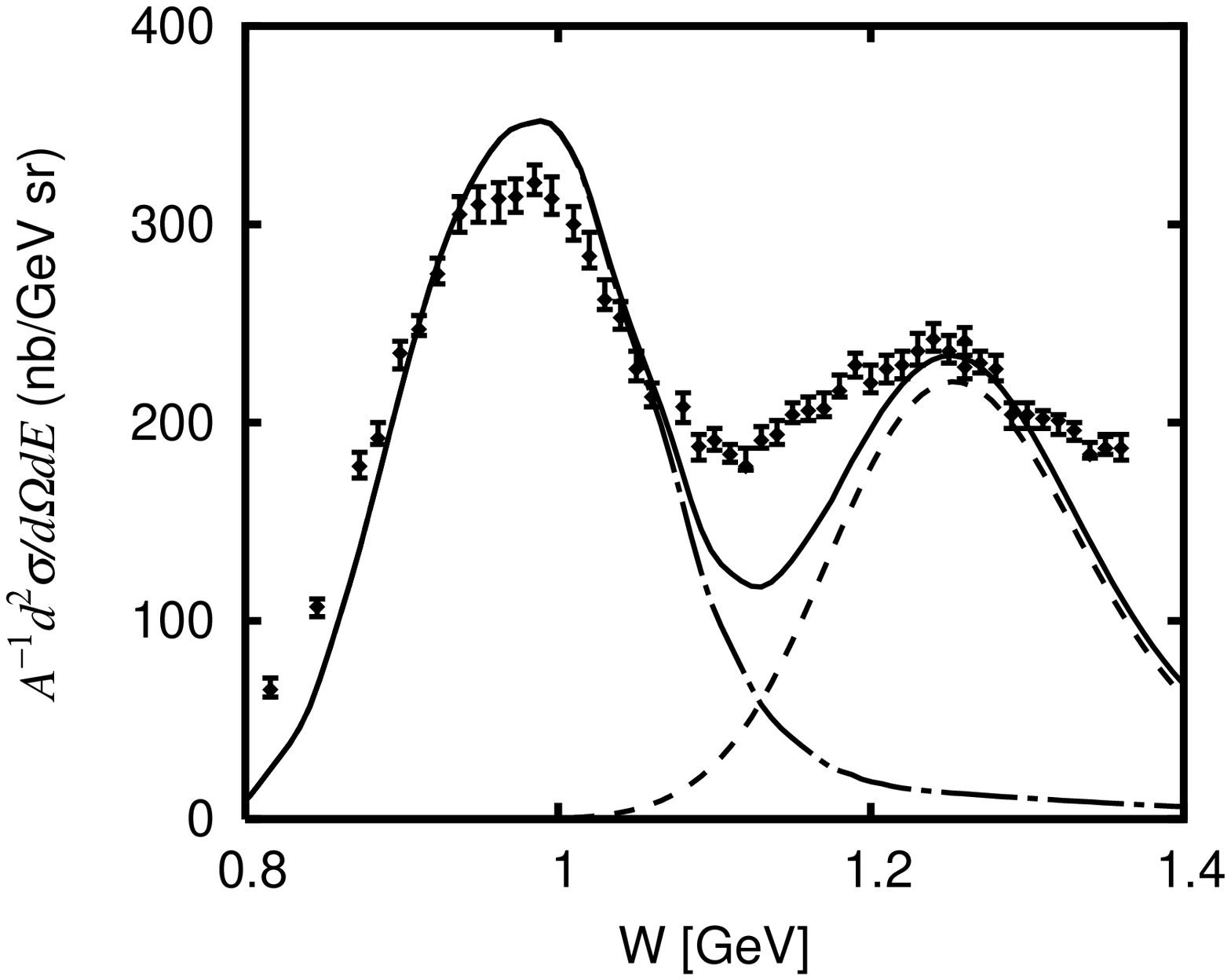}
 \end{minipage}
\caption{\label{fig:c12-1pi} 
(Left) Nuclear effects on differential cross sections for
 $\nu_e + ^{12}{\rm C} \to e^- \pi X$ at $\theta=10^\circ$, normalized
 with the target mass number. 
(Right) Differential cross sections for 
 $e^- + ^{12}{\rm C} \to e^- X$ at $E_e=1.1$~GeV with $\theta_e=37.5^\circ$. 
Data are from Ref.~\cite{data}.
For a description of each curve, see the text. 
}
\end{figure}
Left panel of Fig.~\ref{fig:c12-1pi} shows the nuclear effect on the
differential cross sections for
$\nu_e + ^{12}{\rm C} \to e^- \pi X$, normalized with the mass number,
at the lepton scattering angle $\theta=10^\circ$ as a function of the
final $\pi N$ invariant mass $W$.
The dashed curve shows the differential cross sections for 
the neutrino-induced pion production off the free nucleon, averaged over
the free proton and neutron.
With the Fermi gas effect on the initial nucleon distribution, we obtain
the dashed-double-dotted curve. 
We can see that the Fermi motion broaden the $\Delta$-peak.
By considering the Pauli blocking in addition to the Fermi motion, 
we obtain the dash-dotted curve. 
The Pauli blocking reduces the forward cross sections by about 20\%.
Finally, the solid curve is obtained by replacing the Fermi gas model
with the spectral function taken from Ref.~\cite{benhar}.
The spectral function further broadens the peak, and reduces the height
of it by about 20\%.
For demonstrating the validity of the approach, the model prediction 
for  $e^- + ^{12}{\rm C} \to e^- X$ is
compared with data in Fig.~\ref{fig:c12-1pi} (right).
The QE and 
$\Delta$ peaks reasonably reproduce data.
However, the dip-region is rather underestimated, indicating the need of
going beyond the impulse approximation, and/or more elaborate treatment
of the nuclear correlation.

\section{Coherent $\pi$ production}

The SL model has been applied to the coherent pion production on
$^{12}$C in Ref.~\cite{coh}.
The approach taken in Ref.~\cite{coh} is to combine the elementary
amplitudes from the SL model with the $\Delta$-hole model.
This approach allows us not only to implement the nuclear effects such
as modification of the $\Delta$-propagation and the pion absorption, but
also to describe $\pi$-$A$ scattering, coherent $\pi$ photoproduction and
coherent $\pi$ production in $\nu$-$A$ scattering in a unified manner.
Thus, we can fix parameters relevant to the medium modification of the
$\Delta$-propagation by analyzing $\pi$-$A$ (total and elastic)
scattering, and then we can predict the coherent pion productions.
\begin{figure}[t]
\begin{minipage}[t]{75mm}
   \includegraphics[width=70mm]{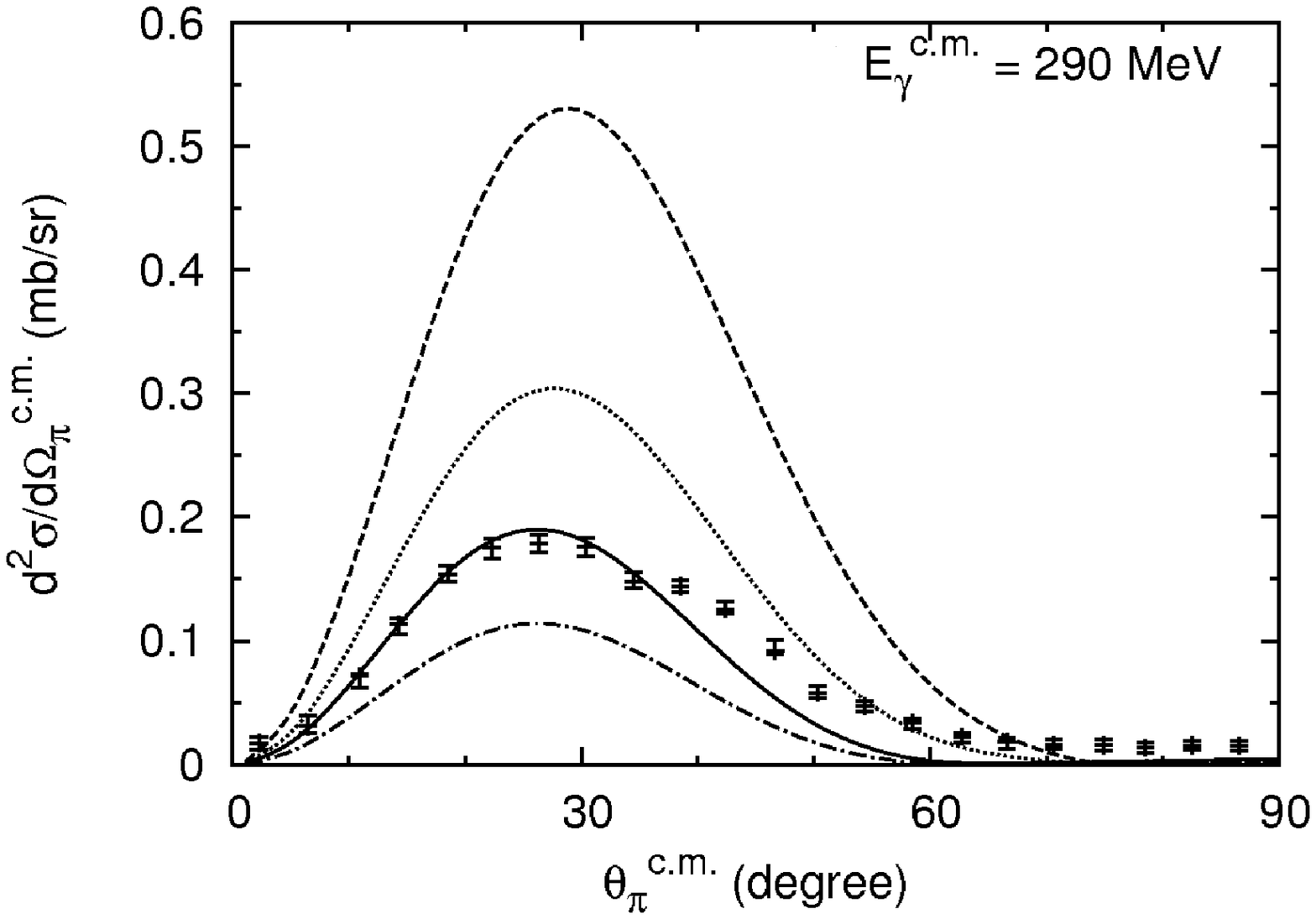}
 \end{minipage}
\begin{minipage}[t]{75mm}
   \includegraphics[width=70mm]{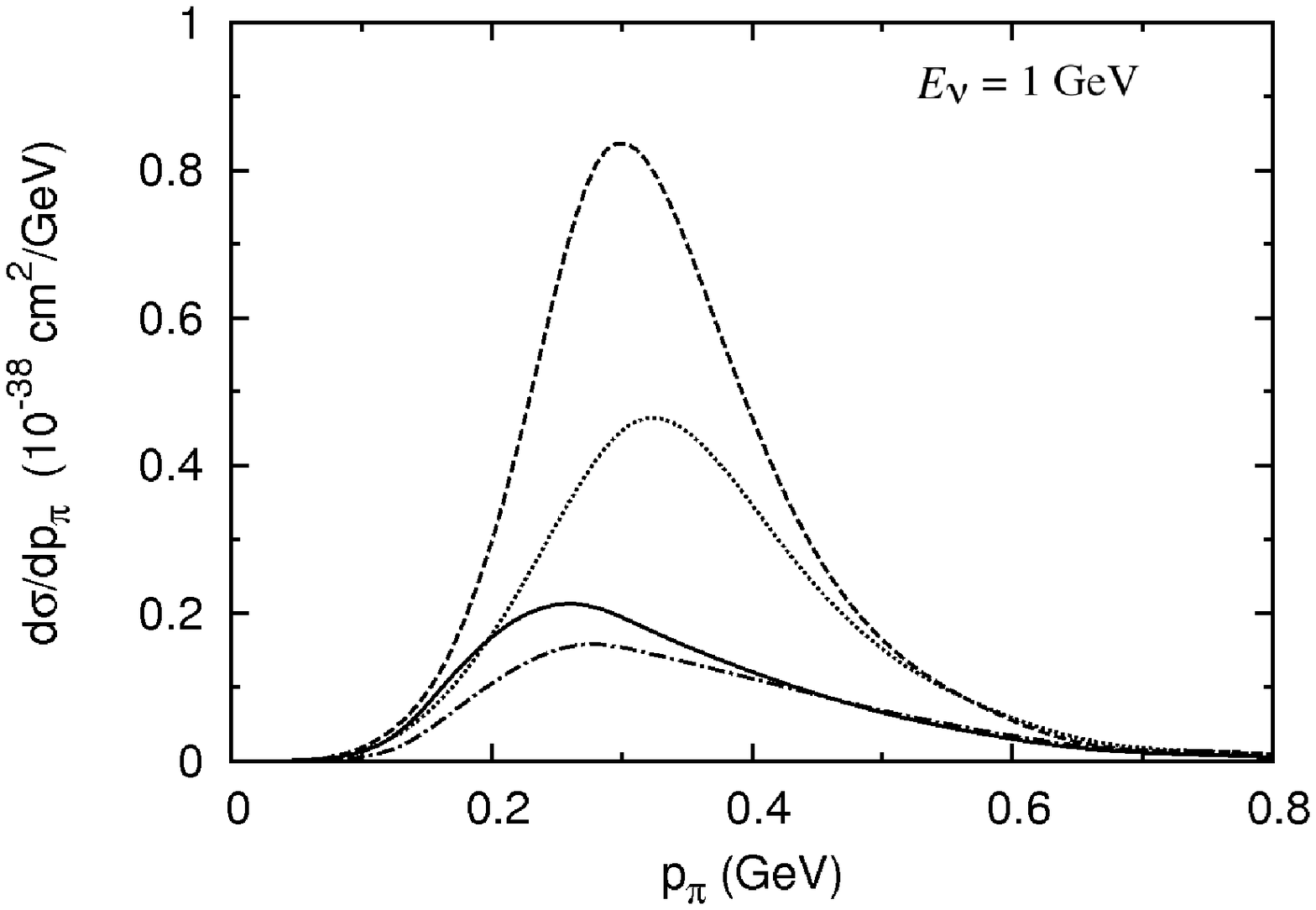}
 \end{minipage}
\caption{\label{fig:coh} (Left) Differential cross sections for the
 coherent pion production off $^{12}$C. Data are from
 Ref.~\cite{krusche}.
(Right) The pion momentum distribution for the charged-current coherent
 pion production in $\nu-^{12}$C scattering. For a description of each
 curve, see the text.
}
\end{figure}

Figure~\ref{fig:coh} shows the nuclear effects and the predictive power
of the model.
The dashed curves in the left and right panels include neither the
medium modification on the $\Delta$-propagation nor the final state
interaction between the pion and nucleus.
The shape of the curves are determined by the elementary amplitude, the
nuclear form factor, and the phase-space factor.
With the medium effect on the $\Delta$, we obtain the dotted curves. 
By further turning on the final state interaction, we obtain the solid
curves. 
The nuclear effects are very significant, and bring the calculation into
a good agreement with data for the photo-production.
This is an important test of the model.

The dash-dotted curves are obtained by turning off the non-resonant amplitudes. 
By observing the difference between the solid and dash-dotted curves, 
we can see a significant contribution from the non-resonant amplitude,
even in the $\Delta$-region.
This is in contrast with the finding in Ref.~\cite{amaro}
that the non-resonant amplitude plays essentially no role.
It is noted that Ref.~\cite{amaro} used a tree-level elementary amplitude, while
Ref.~\cite{coh} used a unitary one from the SL model.
The difference in the reaction mechanism may be responsible for
differences observed in theoretical predictions of the pion momentum
distributions and $E_\nu$-dependence of the total cross sections.

We can average the total cross sections using the neutrino flux from
experiments.
For the charged-current (CC) process, we use the flux from K2K, and obtain
$6.3\times 10^{-40} {\rm cm}^2$ which is consistent with the report from
K2K~\cite{k2k}, $< 7.7\times 10^{-40} {\rm cm}^2$.
For the neutral-current (NC) process, we use the flux from MiniBooNE to obtain
$2.8\times 10^{-40} {\rm cm}^2$ which is still consistent within the rather
large error bar of the preliminary report~\cite{raaf}:
$ 7.7\pm 1.6\pm 3.6\times 10^{-40} {\rm cm}^2$.
However, the CC/NC ratio is not in agreement with the recent report~\cite{kurimoto},
as no theoretical calculations are not.

\section{Future development}

Having seen reasonable descriptions of
pion productions in neutrino(photon, electron)-nucleus scattering 
with the SL model plus nuclear effects,
it is highly hoped to extend this approach to higher mass resonance
region.
This is because 
a model that covers the region from the $\Delta$ to DIS is very useful for
neutrino experiments.
In this energy region, several hadronic channels couple, and $2\pi$
production reactions occupy quite a little portion of final states.
Thus we need a dynamical model that takes care of channel-couplings, and
treat the single and double meson productions on the same footing.
In this context, continuous effort made at the Excited Baryon Analysis
Center (EBAC) in JLab~\cite{ebac} is quite encouraging.
The EBAC has been analyzing world data of 
$\gamma N, \pi N\to \pi N, \pi\pi N, \eta N, KY$ reactions in the
resonance region with a dynamical coupled-channels model (EBAC-DCC model),
and aim to extract resonance information.
The EBAC-DCC model is an extension of the SL model by extending the
coupled-channels from $\pi N$ to 
$\pi N, \eta N, \pi\pi N(\pi\Delta, \sigma N, \rho N), K\Lambda, K\Sigma$, 
and also by including higher resonance states.
It has been demonstrated that the EBAC-DCC model gives a reasonable
description of pion- and photo-induced meson production reactions
from the $\Delta$ to higher mass resonance region~\cite{kamano}.
An extension of the EBAC-DCC model to the weak sector and
neutrino-nucleus reaction, as done with the SL model,
seems a promising future direction.

%%%%%%%%%%%%%%%%%%%%%%%%%%%%%%%%%%%%%%%%%%%%%%%%
%% BACKMATTER
%%%%%%%%%%%%%%%%%%%%%%%%%%%%%%%%%%%%%%%%%%%%%%%%

\begin{theacknowledgments}
The author would like to thank T. Sato, T.-S. H. Lee, 
B. Szczerbinska and K. Kubodera for their collaborations.
This work is supported by the U.S. Department of Energy, Office of
Nuclear Physics Division, under Contract No. DE-AC05-06OR23177
under which Jefferson Science Associates operates Jefferson Lab.
\end{theacknowledgments}



\bibliographystyle{aipproc}   % if natbib is available

\bibliography{sample}

\IfFileExists{\jobname.bbl}{}
 {\typeout{}
  \typeout{******************************************}
  \typeout{** Please run "bibtex \jobname" to optain}
  \typeout{** the bibliography and then re-run LaTeX}
  \typeout{** twice to fix the references!}
  \typeout{******************************************}
  \typeout{}
 }

%%%%%%%%%%%%%%%%%%%%%%%%%%%%%%%%%%%%%%%%%%%
%% The following lines show an example how to produce a bibliography
%% without the help of the BibTeX program. This could be used instead
%% of the above.
%%%%%%%%%%%%%%%%%%%%%%%%%%%%%%%%%%%%%%%%%%%

\end{document}